\documentstyle[12pt]{article}
\title{\baselineskip=9mm
Adiabatic quantum tunneling in heavy-ion sub-barrier fusion \\}
\author{K. Hagino$^{1}$, N. Takigawa$^{1}$,
M. Dasgupta$^{2}$, D.J. Hinde$^{2}$, and J.R. Leigh$^{2}$
\\ \\
\medskip
{\it $^{1}$ Department of Physics,
Tohoku University, Sendai 980--77, Japan}
\\
{\it $^{2}$ Department of Nuclear Physics,} \\
{\it Research School of Physical
Sciences and Engineering, } \\
 {\it Australian National University, Canberra,
ACT 0200, Australia}
}
\date{}

\begin{document}
\baselineskip=9mm

\maketitle

\begin{center}
{\bf Abstract}
\end{center}
          
High precision measurements of the fusion excitation functions for the 
reactions  $^{40}$Ca + $^{194}$Pt, $^{192}$Os clearly demonstrate that 
projectile excitation significantly modifies the potential barrier 
distribution. In sharp contrast, fusion of $^{16}$O + $^{144}$Sm appears 
to show no influence of the projectile excitation on the shape of the 
barrier distribution. These apparently conflicting conclusions are reconciled 
in this work, using realistic coupled--channels calculations, which show 
that high energy states produce an adiabatic potential renormalisation. 
This result indicates that adiabatic effects restrict, in a natural way, 
the states which influence the {\it shape} of a fusion barrier distribution. 
The analysis of barrier distributions thus offers a criterion for the 
relevance of the `counter term' prescription in the Caldeira-Leggett approach.


\newpage

Quantum tunneling plays an important role in a range of diverse phenomena 
in physics and chemistry. Recent attention has been focused on tunneling 
in systems with many degrees of freedom\cite{CL81,JJAP93}. One of the 
interesting aspects of the problem is in determining which of the 
multitude of degrees of freedom must be explicitly included in any 
theoretical description, and which can be omitted. In particular, 
it is essential to define the role of excitation 
energy, or the degree of adiabaticity, in limiting the effectiveness 
of a specific degree of freedom. 

In nuclear physics, heavy-ion fusion reactions at energies near 
and below the Coulomb barrier provide an ideal opportunity to address 
this question. 
In order that fusion reactions occur, the Coulomb barrier 
created by the strong cancellation 
between the repulsive Coulomb force and the  attractive nuclear 
interaction has to be overcome. 
Extensive experimental as well as theoretical studies have revealed 
that couplings of the relative motion of the 
colliding nuclei to nuclear intrinsic degrees of freedom cause 
enhancements of the fusion cross section at subbarrier 
energies, sometimes by several orders of magnitude, over the 
prediction of a single--barrier penetration 
model\cite{B88}. 

In a simple eigenchannel approach, such couplings result in 
the single fusion barrier being replaced by a distribution 
of potential barriers. A method of extracting barrier distributions 
directly from fusion excitation functions was proposed\cite{RSS91}, and 
stimulated precise measurements of the fusion cross sections for 
several systems. In a rigorous theoretical interpretation, the barrier 
distribution representation is valid only if the excitation energy of the 
intrinsic motion is zero\cite{HTBB95}. Nonetheless, this method has been 
successfully applied to analyse data from heavy-ion collisions, where 
each intrinsic degree of freedom carries a finite excitation energy. 
These analyses of the barrier distributions 
have beautifully demonstrated the effects of coupling 
of the relative motion to surface vibrations\cite{MDH94,LDH95}, 
their multi-phonon excitations\cite{SACN95,SACH95}, 
deformations and the associated rotational excitations \cite{LDH95,LLW93} 
of the target nucleus and transfer reactions between the colliding 
nuclei\cite{MDH94,LDH95}. 

Despite these successes, there are apparent conflicts regarding the role of 
projectile excitation. Each barrier distribution for the reactions 
$^{40}$Ca + $^{194}$Pt, $^{192}$Os shows a characteristic structure, with 
a higher energy peak which has been associated with the octupole excitation 
of $^{40}$Ca\cite{BCL96}. 
Calculations of fusion cross-sections for the reactions $^{16}$O + 
$^{154}$Sm, $^{A}$Ge in \cite{GCW94,AKT95} indicated that excitation of 
$^{16}$O is important. In marked contrast, there are no specific features 
in the measured barrier distribution for the $^{16}$O + $^{144}$Sm 
reaction\cite{MDH94} which can be associated with the excitation 
of $^{16}$O; indeed it was shown in Ref. \cite{LDH95} that a good 
theoretical representation of the barrier distribution is destroyed when 
the projectile excitation is included. 

All the above conclusions were based on comparison of the experimental 
results with simplified coupled-channels calculations. The simplification 
has been achieved by using one or more of the following approximations :

\begin{enumerate}

\item the no--Coriolis approximation\cite{HTBBe95}, where the centrifugal 
potential is assumed to be the same for all channels and  equal to that in 
the elastic channel;
\item the linear coupling approximation, where the nuclear coupling 
potential is assumed to be linear {\it w.r.t.} the coordinate of the nuclear 
vibrational excitation; 
\item the constant coupling approximation, where the coupling potential is 
assumed to be constant over the interaction range;
\item intrinsic excitation energies are assumed to be negligible or are 
treated approximately.

\end{enumerate}

The first approximation, common to most coupled channel calculations, 
including those presented in this Letter, has been 
shown to work well for heavy ion fusion calculations\cite{nocor}.  Simplified 
coupled channel calculations\cite{LDH95,SACH95,AKT95,DL87} use the second 
approximation in conjuction with either the third or fourth.
Recent studies\cite{HTDHL96,EsBa97} have shown the linear coupling 
approximation is not valid even in systems where the coupling is weak, and 
that higher order couplings strongly influence the calculated barrier 
distributions. It is therefore probable that in reactions with nuclei like 
$^{16}$O and  $^{40}$Ca, where the couplings to the octupole 
vibrational excitations are strong, barrier distributions calculated
with simplified coupled--channel codes like CCFUS\cite{DL87}  
do not provide a good representation of the fusion process. 

In this Letter we present the results of realistic coupled-channels 
calculations which demonstrate  the effects of non-linear 
coupling and finite excitation energy of  intrinsic nuclear  (environmental) 
degrees of freedom, and resolve the apparently conflicting 
conclusions regarding the influence of the projectile excitation. 
The relevance of the `counter term' 
prescription of Caldeira and Leggett\cite{CL81} 
in heavy-ion fusion reactions is also discussed and 
the double counting problem of coupling effects is clarified.

The coupled channels equations are solved by 
imposing the incoming wave boundary condition to simulate 
the strong absorption inside the fusion barrier. 
The real nuclear potential is assumed to have a Woods-Saxon shape 
and the depth was chosen to reproduce the experimental 
fusion cross sections at high energies using 
the single--barrier penetration  model. 
The values of deformation parameters are extracted from the 
reduced transition probabilities using 
$\beta_{\lambda}=\frac{4\pi}{3ZR^{\lambda}}
\sqrt{\frac{B(E\lambda)\uparrow}{e^2}} $. 
The parameters of the calculations are listed in Table~1. 

In order to show the inadequacies of the often used linear coupling 
approximation, calculations were performed for the $^{16}$O + $^{144}$Sm 
reaction using the linear coupling approximation. 
The results of our calculations for the fusion excitation function and 
the barrier distribution are shown in Fig.~1. In the following discussion we 
concentrate on the latter since they are a more sensitive way to compare 
experimental data and calculations. 
The dotted line shows the result when the excitation of $^{16}$O 
is not included in the calculations. This calculation 
well reproduces the features of the experimental barrier distribution. 
Calculations including  the excitation of the lowest-lying 
octuple state of $^{16}$O are shown by the solid line. Even though 
the experimental barrier distribution around 
the lower energy peak ($\sim$ 60 MeV) is reproduced, significant strength is 
missing around the higher energy peak near 65 MeV. 
A similar discrepancy between theory and experimental data 
was encountered in Ref.~\cite{LDH95}, where calculations, shown by the 
long--dashed line, were performed using a modified version of the CCFUS 
code. It is clear that both calculations which include the 
octupole excitation of $^{16}$O in the linear coupling approximation  
fail to reproduce the experimental barrier distribution.  

Realistic coupled channels calculations were then performed, where  
the couplings to the octupole vibrations of both $^{16}$O and $^{144}$Sm 
are treated to all orders; {\it{i.e.}} the nuclear interaction is not 
expanded with respect to the deformation parameter, and the coupling matrix 
elements are numerically evaluated at each internuclear separation 
\cite{HTDHL96}. It is remarkable that these calculations, shown in Fig.~2, 
re-establish the double--peaked structure seen in the experimental data, 
which was absent in the equivalent linear coupling calculations. 
The shape of the barrier distribution obtained by 
including the octupole vibration of $^{16}$O using  all order coupling 
is now very similar to that obtained by ignoring it. 
This similarity becomes particularly evident when the calculated 
barrier distribution is shifted in energy, as shown 
by the dashed line in the figure. 
This is consistent with the general conclusion that the main effect of 
the coupling to inelastic channels whose excitation energies are larger   
than the curvature of the bare fusion barrier, i.e. an adiabatic 
coupling, is to introduce a static potential shift\cite{THAB94,THA95}, 
and hence, the shape of the barrier distribution does not change 
unless the coupling form factor itself has a strong radial dependence. 

In macroscopic quantum tunneling in 
condensed matter physics, the so--called 
counter term is often introduced in order to compensate for 
the static potential 
renormalization due to the coupling to the environment\cite{CL81}. 
In contrast, in heavy-ion reactions, 
one usually estimates the bare potential, for example 
by fitting the fusion cross section at high energies, and discusses 
the effects of channel coupling without introducing the counter term. 
Fig. 2 shows that this approach reproduces the experimental fusion 
cross sections and fusion barrier distributions without explicitly 
taking into account the excitation of the octupole vibrational state 
of $^{16}$O. This indicates that the effects of its excitation is already 
included in the bare potential. If this is the case, the effect 
of the coupling to the 3$^-$ state 
of $^{16}$O is double counted if the coupled channels 
calculations explicitly take it into account, resulting in a 
dramatic  overestimate of the  the experimental cross-sections. 
A recipe to cure this problem is to introduce the counter term 
as in condensed matter physics. Since the experimental data are well 
reproduced when the calculated distributions are shifted to higher energies 
by 2 MeV, this shift evidently mimics the effects of the counter term. 

In contrast to the $^{16}$O+$^{144}$Sm case, the analyses of 
$^{40}$Ca + $^{192}$Os and $^{194}$Pt reactions, also performed 
using simplified coupled channel calculations\cite{BCL96}, 
suggest that the excitation of 
$^{40}$Ca is important in determining the observed barrier distribution. 
An important difference between the $^{16}$O and 
$^{40}$Ca projectiles is that 
the excitation energy of the octupole vibration in the latter 
is smaller and nearly equal to the energy scale of the curvature of the 
fusion barrier, hence the coupling is intermediate between adiabatic 
and sudden. It is therefore interesting to investigate the degree of 
adiabaticity of the octupole excitation of the $^{40}$Ca projectile. 

The results of the coupled channels calculations are compared with 
the experimental data in Fig. 3. All order couplings to 
both the target and the projectile excitations have been included. 
Although $^{194}$Pt and $^{192}$Os are transitional nuclei 
which lie between the $\gamma$-unstable and rotational limits 
in the interacting boson model\cite{LK92}, we have assumed that they are 
rigid rotors with axial symmetry. The ground state rotational band of the 
target nucleus, with states up to the 10$^+$ member, has been included 
in the calculations. When the  $^{40}$Ca  excitation is ignored, barrier 
distributions are obtained which are similar to those expected for 
a classically deformed nucleus and these are inconsistent with the 
experimental data. When the octupole excitation of $^{40}$Ca is included 
in the calculations,  a higher energy peak is introduced which agrees 
well with that observed in each reaction. The mutual excitation channels 
up to 4$^+\bigotimes 3^-$, the former and the latter refering to the 
targets and the projectile respectively, are also included in the 
calculations. It is apparent that the projectile excitation significantly 
affects the shape of the barrier distribution in this case, as suggested 
in the simplified coupled channel calculations in Ref.~\cite{BCL96}. 

As has been shown in the discussions for $^{16}$O + $^{144}$Sm reactions, 
the correct treatment of the coupling, without making the linear coupling 
approximation, significantly 
reduces the effect of projectile excitation on the shape of the barrier 
distribution. Calculations of the CCFUS-type, which fail in these regards, 
would therefore be expected to predict larger coupling effects than observed 
experimentally. The apparent success of the CCFUS calculations reported in 
Ref. \cite{BCL96} was probably due to the compensation for this overestimate 
by the use of a smaller deformation parameter than that obtained from the 
octupole transition strength.

The theoretical calculations for the reactions with the $^{40}$Ca projectile 
still significantly underestimate the fusion cross section at low energies, 
even after the excitation of the projectile is taken into account. 
As suggested in Ref.~\cite{BCL96}, coupling to 
transfer channels, which have been ignored in the present calculations, 
might enhance the fusion cross section at low energies. 

In summary, we have performed coupled-channels calculations for the fusion 
reactions $^{16}$O + $^{144}$Sm and $^{40}$Ca + $^{194}$Pt, $^{192}$Os. 
The calculations with  full order coupling 
show that the dominant effect of the excitation of the 
$^{16}$O octupole state at 6.1 MeV is to renormalise the static 
potential without significantly 
changing the shape of the barrier distribution. On the other hand, the 
excitation of the 3$^-$ state at 3.7 MeV in $^{40}$Ca 
introduces well defined peaks 
in the barrier distribution. These results suggest a natural 
limit to the energy of states which need to be considered explicitly 
in coupled-channels calculations. The myriad of weak, high energy 
excitations which might be possible, contribute only to a potential 
renormalisation without affecting the shape of the barrier distribution.  
The effects of these excitations can then be included in the  bare potential 
in coupled channels calculations. If these channels are explicitly included 
in the coupled channels calculations without introducing the counter term, 
they could be double counted depending on the choice of the bare potential. 

It has been shown that  in order to interpret the high precision fusion 
excitation functions, that have recently become available, it is vital to 
perform exact coupled channel calculations which treat the excitation energy 
and the radial dependence of the coupling form factor correctly. 
Whilst CCFUS-based calculations have apparently been very successful in 
reproducing observed barrier distributions, it is clear from our results that 
care must be taken in their interpretation; the approximations used are 
unreliable even for relatively weak coupling strengths. Exact 
coupled-channels calculation is the only reliable means of quantitatively 
understanding the barrier distributions. 

\medskip

The authors thank R. Vandenbosch for providing them with the 
experimental data and S. Kuyucak for useful discussions.
K.H. and N.T. also thank the Australian National University for its 
hospitality and for partial support for this project.
The work of K.H. was supported by the Japan Society for the Promotion 
of Science for Young Scientists.
This work was supported by the Grant-in-Aid for General
Scientific Research,
Contract No.06640368 and No.08640380, and the Grant-in-Aid for Scientific
Research on Priority Areas, Contract No.05243102 and 08240204  
from the Japanese Ministry of Education, Science and Culture, 
and a bilateral program of JSPS between Japan and Australia. 

\newpage
\begin{table}
\caption{ Parameters used in the coupled channels calculations for the 
indicated reactions. }
\begin{tabular}{|l|lcccc|ccc|}
\hline
 &\multicolumn{5}{|c|}{Channel Couplings}&\multicolumn{3}{|c|}{Potential 
parameters}\\
\multicolumn{1}{|c|}{Reaction} &\multicolumn{1}{c}{ Nucleus}& 
\multicolumn{1}{c}{ Type}& 
\multicolumn{1}{c}{$\lambda^{\pi}$}&\multicolumn{1}{c}{ E$^*$ (MeV)}&
\multicolumn{1}{c}{$\beta_{\lambda}$}&\multicolumn{1}{|c}{ V (MeV)}&
\multicolumn{1}{c}{ r$_0$ (fm)} & \multicolumn{1}{c|}{a (fm)}\\
\hline
 & & & & & & & & \\
$^{16}$O + $^{144}$Sm &  $^{144}$Sm& vib& 3$^{-}$ & 
1.81 & 0.205& 105.1 & 1.1 & 0.75\\
 &\ $^{16}$O & vib& 3$^{-}$ & 6.13 & 0.733 &  & & \\
 & & & & & & & & \\
 $^{40}$Ca + $^{194}$Pt 
&\  $^{40}$Ca & vib& 3$^{-}$& 3.70 & 0.339 & 330.0 & 1.0 & 0.84\\
&$^{194}$Pt& rot& 2$^{+}$ 
& 0.328& $\beta_2=-$0.154 & & & \\
  & & & &  & $\beta_4=-$0.045 &  & &  \\
 & & & & & & & & \\
 $^{40}$Ca + $^{192}$Os &$^{192}$Os& rot& 2$^{+}$ 
& 0.206& $\beta_2=$0.167 & \ 148.0 & 1.1 & 0.84\\
  & & & &  & $\beta_4=-$0.043 &  & &  \\
 \hline
\end{tabular}
\label{tab1}
\end{table}

\begin{center}
{\bf Figure Captions}
\end{center}

\noindent
{\bf Fig.1:}
Fusion excitation functions (upper 
panel) and the barrier distributions (lower panel) for the 
$^{16}$O + $^{144}$Sm reaction. 
The experimental data (filled circles) are taken from Ref.~\cite{LDH95}. 
The linear coupling approximation is used 
in the coupled channels calculations. 
In all calculations, the effects of the octupole vibration of 
$^{144}$Sm are taken into account.
The dotted line is the results when $^{16}$O is treated as inert. 
The solid line is the result of the coupled channels calculations when 
the coupling to the octupole vibration of $^{16}$O is also 
taken into account; the dashed line is the result of an
 equivalent CCFUS calculation. 

\noindent
{\bf Fig.2:} As Fig.1, but for the case when the 
coupled-channels calculations have been performed by including 
all order coupling.  
The meaning of the solid and the dotted lines is the same as in Fig. 1, 
while the dashed line is the same calculation as the solid line with 
the average barrier increased by 2 MeV.

\noindent
{\bf Fig.3:}
The comparison of the experimental fusion cross sections 
(upper panels) and fusion barrier distributions (lower 
panels) for the $^{40}$Ca + $^{194}$Pt, $^{192}$Os reactions with 
the coupled channels calculations. 
In all calculations, the effects of the excitation of 
the target nuclei are treated in the rotational model and 
 couplings to all orders are included. 
The dotted lines are the results when $^{40}$Ca 
is treated as inert. The solid lines include the coupling to 
the octupole vibrational state in $^{40}$Ca.


\newpage

\begin{thebibliography}{99} 

\bibitem{CL81}A.O. Caldeira and A.J. Leggett, Phys. Rev. Lett. 
{\bf 46}, 211(1981).

\bibitem{JJAP93}{\it Proceedings of the Fourth International Symposium on
Foundations of Quantum Mechanics,} edited by M. Tsukada {\it et al.,}
Japanese Journal of Applied Physics Series Vol. 9
( Publication Office of Japanese Journal of Applied Physics,
Tokyo, 1993).

\bibitem{B88}M. Beckerman, Rep. Prog. Phys. {\bf 51}, 1047(1988); 
A.B. Balantekin and N. Takigawa, Rev. Mod. Phys. (to be published). 

\bibitem{RSS91}N. Rowley et al,
Phys. Lett. {\bf B254}, 25(1991).

\bibitem{HTBB95}K. Hagino et al,
Phys. Rev. C{\bf 51}, 3190(1995).
 
\bibitem{MDH94}C.R. Morton et al,
Phys. Rev. Lett. {\bf 72}, 4074(1994).

\bibitem{LDH95}J.R. Leigh et al,
Phys. Rev. C{\bf 52}, 3151(1995).

\bibitem{SACN95}A.M. Stefanini et al,
Phys. Rev. Lett. {\bf 74}, 864(1995).
 
\bibitem{SACH95}A.M. Stefanini et al,
Phys. Rev. C{\bf 52}, R1727(1995).

\bibitem{LLW93}R.C. Lemmon et al,
Phys. Lett. {\bf B316}, 32(1993).

\bibitem{BCL96}J.D. Bierman et al,
Phys. Rev. Lett. {\bf 76},
1587(1996); Phys. Rev. C{\bf 54}, 3068(1996). 

\bibitem{GCW94}P.R.S. Gomes et al,
Phys. Rev. C{\bf 49}, 245(1994).
  
\bibitem{AKT95}E.F. Aguilera et al,
Phys. Rev. C{\bf 52}, 3103(1995).

\bibitem{HTBBe95}K. Hagino et al,
Phys. Rev. C{\bf 52}, 286(1995); 
N. Takigawa and K. Ikeda, 
in {\it Proceedings of the Symposium on Many Facets of Heavy Ion 
Fusion Reactions}, edited by W. Henning {\it et al.}(Argonne 
National Laboratory Report No. ANL-PHY-87-1), 1986, p.613.

\bibitem{nocor} H. Esbensen et al, Phys. Rev. C{\bf 36}, 2359(1987). 

\bibitem{DL87}C.H. Dasso  and S. Landowne, Comp. Phys. Commun.
{\bf 46}, 187(1987); J.O. Fern\'{a}ndez Niello et al,
{\it ibid.}{\bf 54}, 409 (1989).
 
\bibitem{HTDHL96}K. Hagino et al,
Phys. Rev. C, in press.

\bibitem{EsBa97}H. Esbensen and B.B. Back, Phys. Rev. C{\bf 54}, 
3109(1996).

\bibitem{THAB94}N. Takigawa et al,
Phys. Rev. C{\bf 49}, 2630(1994).

\bibitem{THA95}N. Takigawa et al,
Phys. Rev. C{\bf 51},
187(1995).

\bibitem{LK92}
V.-S. Lac and S. Kuyucak, Nucl. Phys. {\bf A539}, 418 (1992).

\end{thebibliography}
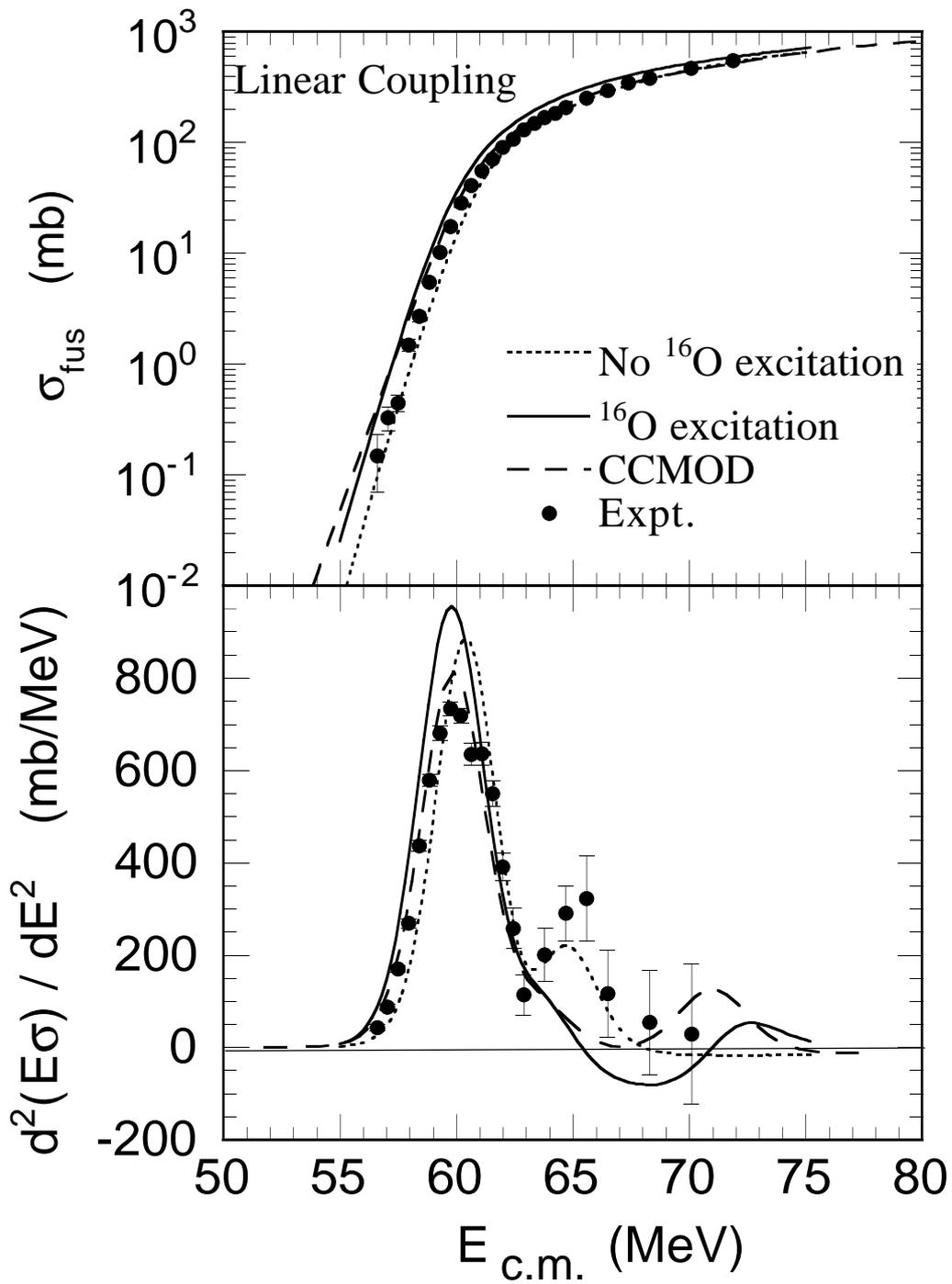
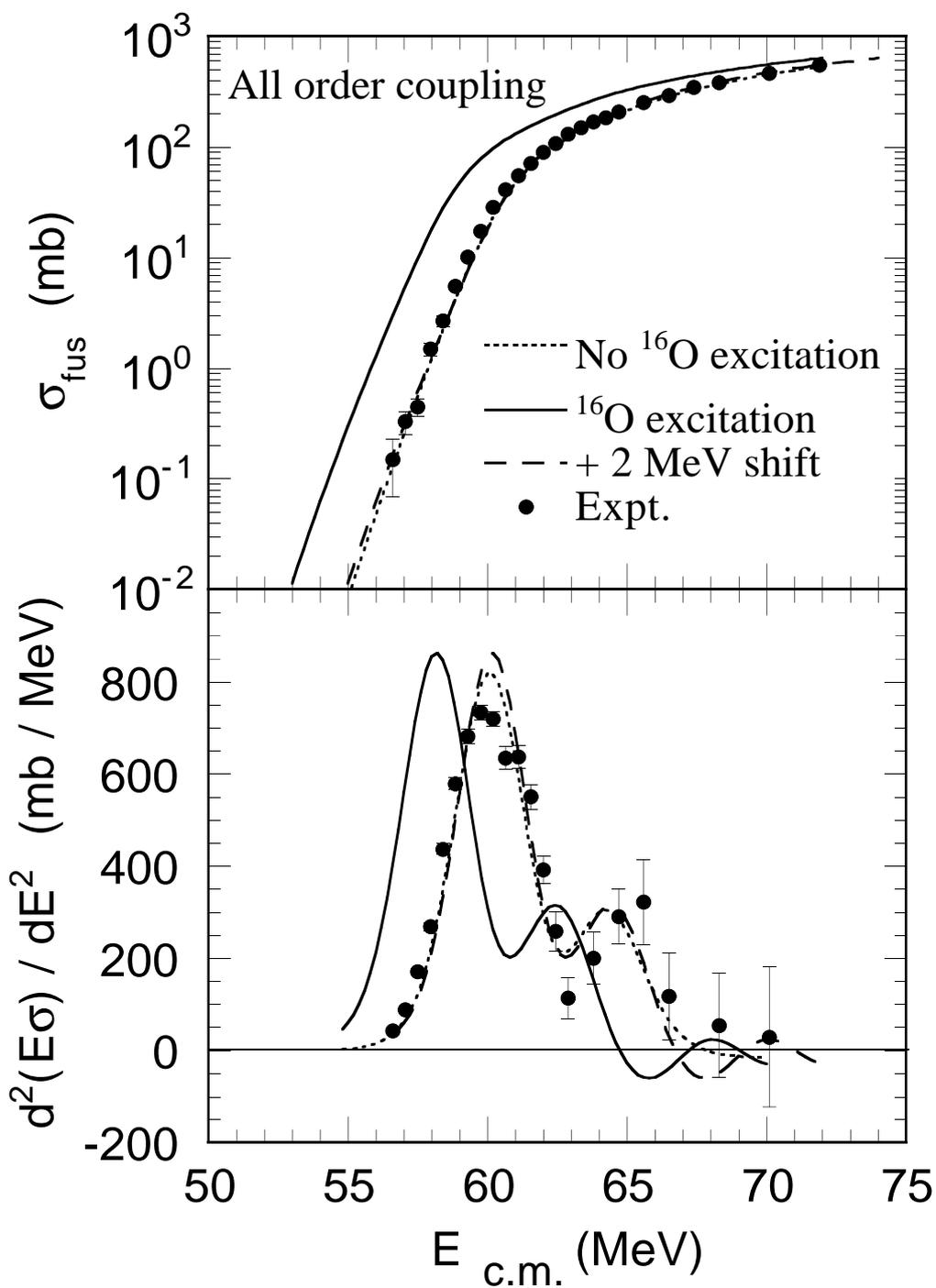
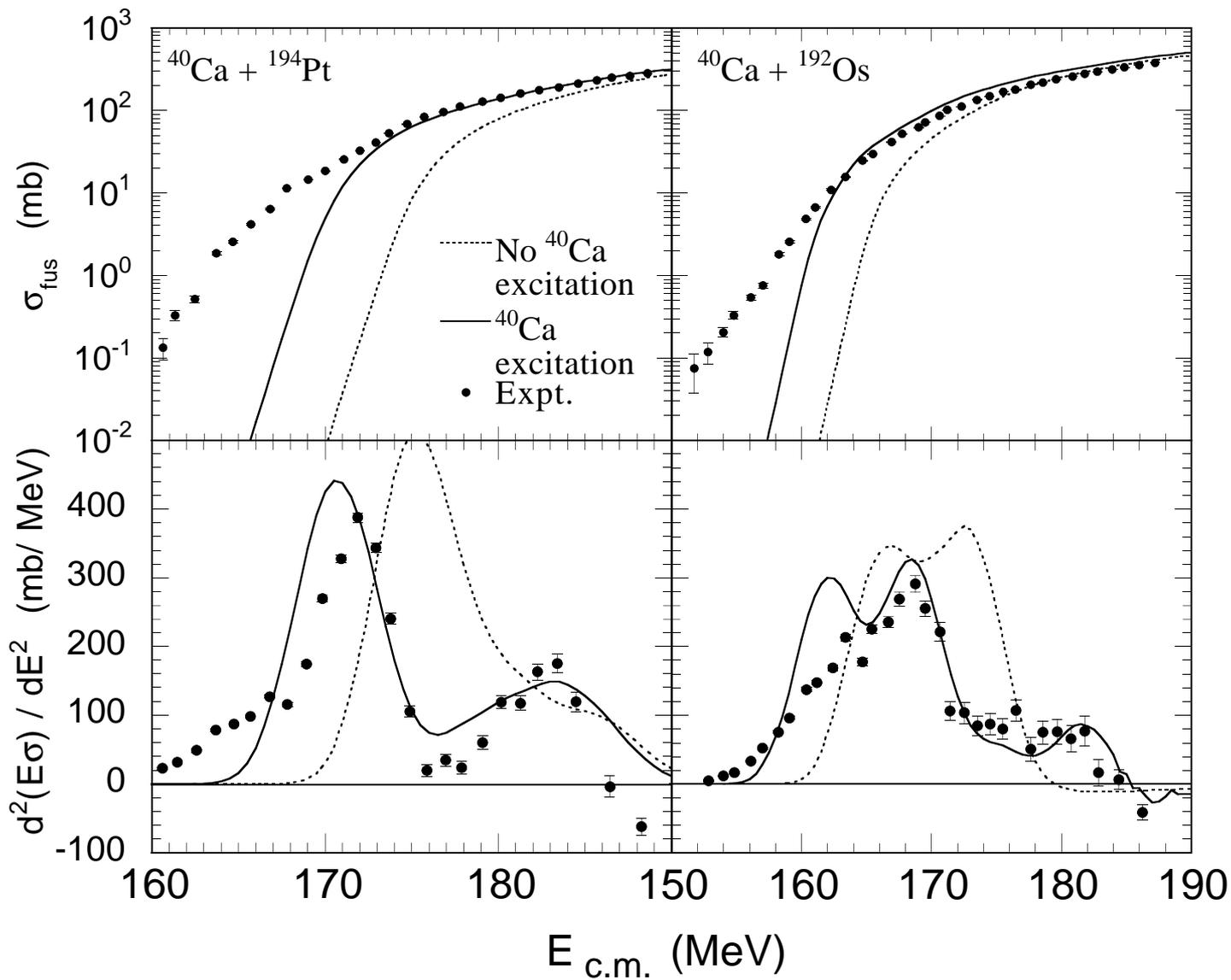
\end{document}